\documentclass[12pt]{article}
\usepackage{a4,graphicx,cite}
\usepackage[small,bf]{caption}

\usepackage{color}

\newcommand{\sixth}{\mbox{\small $\frac{1}{6}$}}         
\newcommand{\half}{\mbox{\small $\frac{1}{2}$}}          
\newcommand{\third}{\mbox{\small $\frac{1}{3}$}}         
\newcommand{\twothird}{\mbox{\small $\frac{2}{3}$}}      
\newcommand{\imp}{\mbox{\tiny $I\!M\!P$}}                
\newcommand{\R}{\mbox{\tiny $R$}}                        
\newcommand{\So}{\mbox{\tiny $S$}}                       

\def\lsim{\mathrel{\rlap{\lower4pt\hbox{\hskip1pt$\sim$}}
    \raise1pt\hbox{$<$}}}                
\def\gsim{\mathrel{\rlap{\lower4pt\hbox{\hskip1pt$\sim$}}
    \raise1pt\hbox{$>$}}}                

\begin{document}

\title{
\vspace{-3.25cm}
\flushright{\small ADP-16-46/T1002} \\
\vspace{-0.35cm}
{\small DESY 16-241} \\
\vspace{-0.35cm}
{\small Edinburgh 2016/19} \\
\vspace{-0.35cm}
{\small Liverpool LTH 1116} \\
\vspace{-0.35cm}
{\small December 14, 2016}  \\
\vspace{0.5cm}
\centering{\Large \bf Flavour breaking effects in the pseudoscalar
          meson decay constants}}

\author{\large
         V.~G. Bornyakov$^a$, R. Horsley$^b$,
         Y. Nakamura$^c$, H. Perlt$^d$, \\ 
         D. Pleiter$^e$, P.~E.~L. Rakow$^f$, 
         G. Schierholz$^g$, A. Schiller$^d$, \\ 
         H. St\"uben$^h$ and J.~M. Zanotti$^i$ \\[1em]
         -- QCDSF-UKQCD Collaborations -- \\[1em]
        \small $^a$ Institute for High Energy Physics, 
               Protvino, \\[-0.5em]
        \small 142281 Protvino, Russia, \\[-0.5em]
        \small Institute of Theoretical and Experimental Physics,
               Moscow, \\[-0.5em]
        \small 117259 Moscow, Russia, \\[-0.5em]
        \small School of Biomedicine, 
               Far Eastern Federal University, \\[-0.5em]
        \small 690950 Vladivostok, Russia      \\[0.25em]
        \small $^b$ School of Physics and Astronomy,
               University of Edinburgh, \\[-0.5em]
        \small Edinburgh EH9 3FD, UK \\[0.25em]
        \small $^c$ RIKEN Advanced Institute for
               Computational Science, \\[-0.5em]
        \small Kobe, Hyogo 650-0047, Japan \\[0.25em]
        \small $^d$ Institut f\"ur Theoretische Physik,
               Universit\"at Leipzig, \\[-0.5em]
        \small 04109 Leipzig, Germany \\[0.25em]
        \small $^e$ J\"ulich Supercomputing Centre,
               Forschungszentrum J\"ulich, \\[-0.5em]
        \small 52425 J\"ulich, Germany, \\[-0.5em]
        \small Institut f\"ur Theoretische Physik,
               Universit\"at Regensburg, \\[-0.5em]
        \small 93040 Regensburg, Germany \\[0.25em]
        \small $^f$ Theoretical Physics Division,
               Department of Mathematical Sciences, \\[-0.5em]
        \small University of Liverpool,
               Liverpool L69 3BX, UK \\[0.25em]
        \small $^g$ Deutsches Elektronen-Synchrotron DESY, \\[-0.5em]
        \small 22603 Hamburg, Germany \\[0.25em]
        \small $^h$ Regionales Rechenzentrum, Universit\"at Hamburg, \\[-0.5em]
        \small 20146 Hamburg, Germany \\[0.25em]
        \small $^i$ CSSM, Department of Physics,
               University of Adelaide, \\[-0.5em]
        \small Adelaide SA 5005, Australia}

\date{}

\maketitle



\begin{abstract}
   The SU(3) flavour symmetry breaking expansion in up, down and
   strange quark masses is extended from hadron masses to meson
   decay constants. This allows a determination of the ratio of
   kaon to pion decay constants in QCD. Furthermore when using
   partially quenched valence quarks the expansion is such that
   SU(2) isospin breaking effects can also be determined. It is
   found that the lowest order SU(3) flavour symmetry breaking
   expansion (or Gell-Mann--Okubo expansion) works very well.
   Simulations are performed for 2+1 flavours of clover fermions
   at four lattice spacings.
\end{abstract}







\section{Introduction}
\label{introduction}


One approach to determine the ratio $|V_{us}/V_{ud}|$ of 
Cabibbo–-Kobayashi–-Maskawa (CKM) matrix elements, as suggested in
\cite{marciano04a}, is by using the ratio of the experimentally
determined pion and kaon leptonic decay rates
\begin{eqnarray}
   {\Gamma(K^+ \to \mu^+ \nu_\mu) \over
                             \Gamma(\pi^+ \to \mu^+ \nu_\mu)}
      = \left|{V_{us} \over V_{ud}}\right|^2 \,
        \left( f_{K^+} \over f_{\pi^+} \right)^2 \,
        { M_{K^+} \over M_{\pi^+} } \,
        \left( 1 - m_\mu^2/M^2_{K^+} \over 1 - m_\mu^2/M^2_{\pi^+} \right)^2 \,
        ( 1 + \delta_{\rm em} ) \,,
\end{eqnarray}
(where $M_{K^+}$, $M_{\pi^+}$ and $m_\mu$ are the particle masses,
and $\delta_{\rm em}$ is an electromagnetic correction factor).
This in turn requires the determination of the ratio of kaon to pion
decays constants, $f_{K^+} / f_{\pi^+}$, a non-perturbative task, 
where the lattice approach to QCD may be of help. For some recent
work see, for example,
\cite{durr10a,aoki10a,engel11a,bazavov13a,dowdall13a,bazavov14a,
blum14a,carrasco14a,durr16a}.

The QCD interaction is flavour-blind and so when neglecting electromagnetic
and weak interactions, the only difference between the quark flavours
comes from the mass matrix. In this article we want to examine how this
constrains meson decay matrix elements once full $SU(3)$ flavour
symmetry is broken, using the same methods as we used in 
\cite{bietenholz10a,bietenholz11a} for hadron masses. In particular
we shall consider pseudoscalar decay matrix elements and give
an estimation for $f_K/f_\pi$ and $f_{K^+}/f_{\pi^+}$ (ignoring
electromagnetic contributions).


\section{Approach}
\label{approach}

\begin{figure}[h]
\begin{center}

\begin{minipage}{0.45\textwidth}

   \begin{center}
      \includegraphics[width=5.00cm]{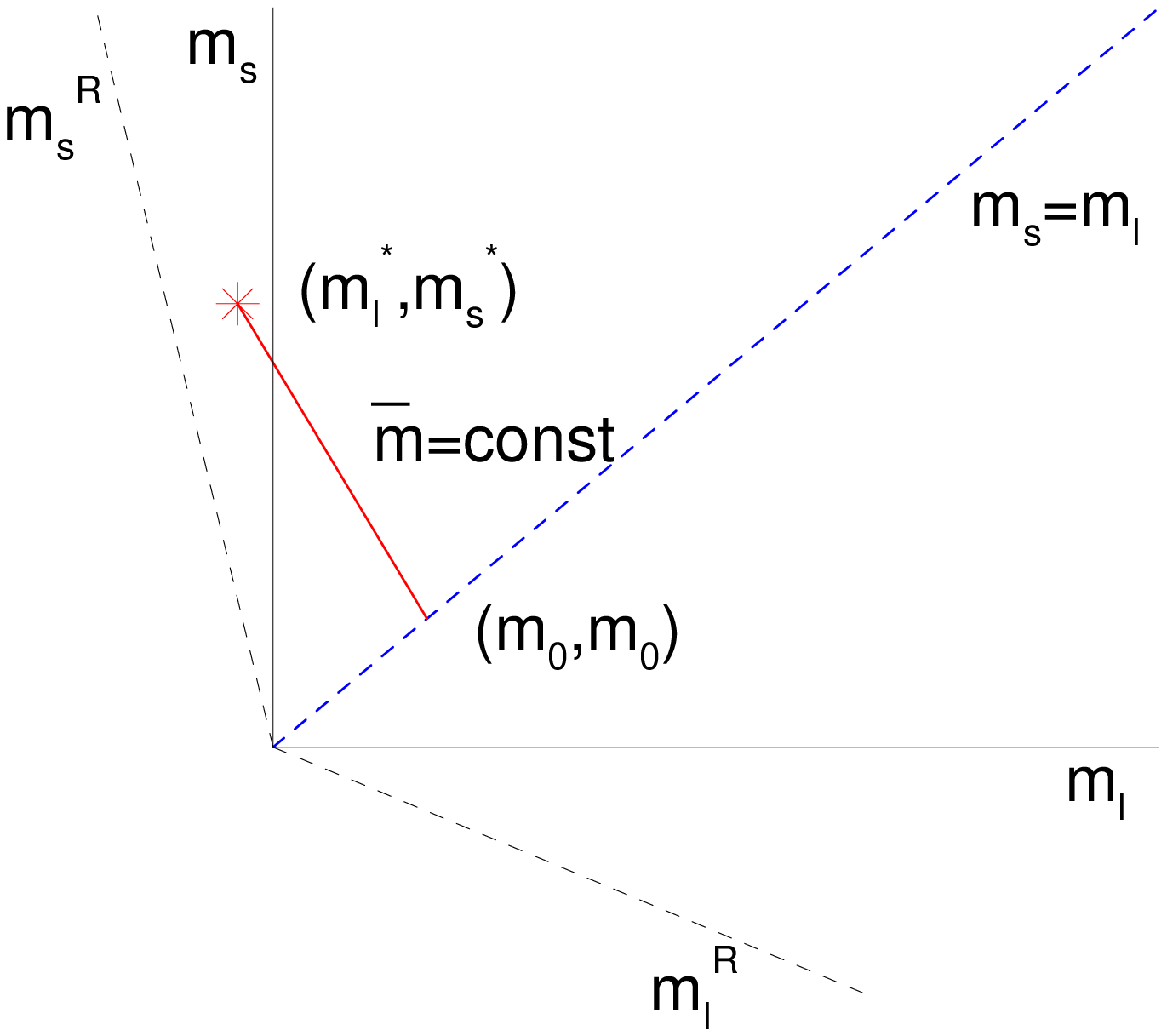}
   \end{center} 

\end{minipage}\hspace*{0.05\textwidth}
\begin{minipage}{0.45\textwidth}

   \begin{center}
      \includegraphics[width=5.00cm]{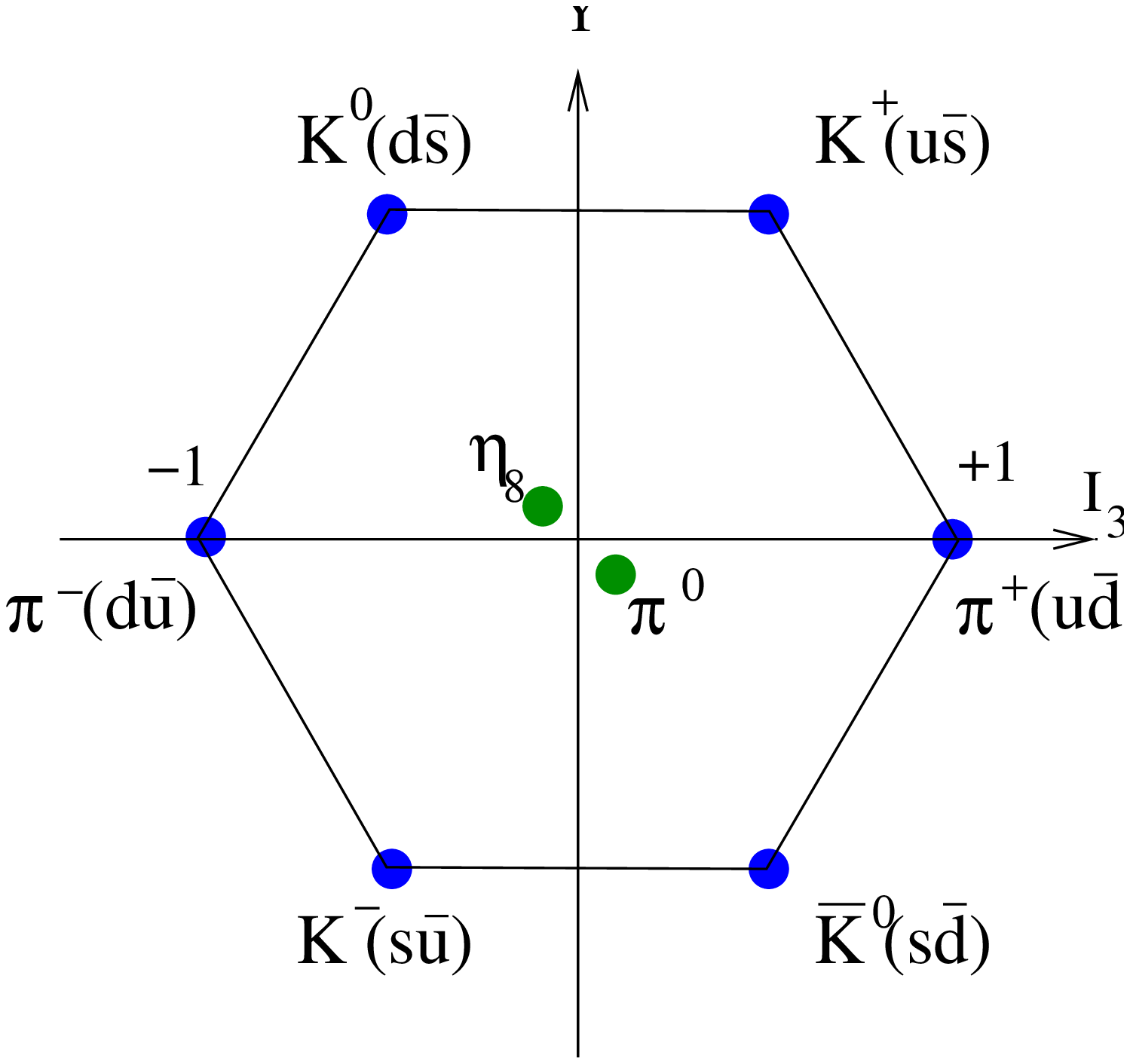}
   \end{center} 

\end{minipage}

\end{center}
\caption{LH panel: Sketch of the path for the case of two mass
         degenerate quarks, $m_u = m_d \equiv m_l$, from a point
         on the $SU(3)$ flavour symmetric line $(m_0,m_0)$ to
         the physical point denoted with a $^*$: $(m_l^*, m_s^*)$.
         RH panel: The pseudoscalar octet meson.}
\label{path+octet}
\end{figure}

In lattice simulations with three dynamical quarks
there are many paths to approach the physical point where the
quark masses take their physical values. The choice adopted here is
to extrapolate from a point on the $SU(3)$ flavour symmetry line
keeping the singlet quark mass $\overline{m}$ constant,
as illustrated in the left panel of Fig.~\ref{path+octet},
for the case of two mass degenerate quarks $m_u = m_d \equiv m_l$.
This allows the development of an $SU(3)$ flavour symmetry
breaking expansion for hadron masses and matrix elements, i.e.\
an expansion in
\begin{eqnarray}
   \delta m_q  = m_q - \overline{m} \,, \quad \mbox{with} \quad
            \overline{m} = \third(m_u + m_d + m_s) \,,
\end{eqnarray}
(where numerically $\overline{m} = m_0$). From this definition we have
the trivial constraint 
\begin{eqnarray}
   \delta m_u+\delta m_d+\delta m_s = 0 \,.
\label{trivial_constraint}
\end{eqnarray}
The path to the physical quark masses is called the
`unitary line' as we expand in the same masses for the sea and
valence quarks. Note also that the expansion coefficients are
functions of $\overline{m}$ only, which provided we keep
$\overline{m} = \mbox{const.}$ reduces the number of allowed
expansion coefficients considerably.

As an example of an $SU(3)$ flavour symmetry breaking expansion,
\cite{bietenholz11a}, we consider the pseudoscalar masses, and
find to next-to-leading-order, NLO, (i.e.\ $O((\delta m_q)^2)$).
\begin{eqnarray}
   M^2(a\overline{b})
      &=& M^2_0 + \alpha(\delta m_a + \delta m_b)
                                                         \nonumber  \\
      & & \phantom{M^2_0}
          + \beta_0\sixth(\delta m_u^2 + \delta m_d^2 + \delta m_s^2)
          + \beta_1(\delta m_a^2 + \delta m_b^2)
          + \beta_2(\delta m_a - \delta m_b)^2
                                                         \nonumber  \\
      & & \phantom{M^2_{0\pi}}
          + \ldots \,,
\label{Mpi2_NLO_expan}
\end{eqnarray}
where $m_a$, $m_b$ are quark masses with $a, b = u, d, s$. This
describes the physical outer ring of the pseudoscalar meson octet
(the right panel of Fig.~\ref{path+octet}). Numerically we can also
in addition consider a fictitious particle, where $a = b = s$, 
which we call $\eta_s$. We have further extended the expansion to 
the next-to-next-to-leading or NNLO case, \cite{horsley12a}.
As the expressions start to become unwieldy, they have been relegated
to Appendix~\ref{NNLO_expan}. (Octet baryons also have equivalent
expansions,  \cite{horsley12a}.)

The vacuum is a flavour singlet, so meson to vacuum matrix elements
$\langle 0 | \widehat{\cal O} | M \rangle$
are proportional to $1 \otimes 8 \otimes 8$ tensors, i.e.\ 
$8 \otimes 8$ matrices, where $\widehat{\cal O}$ is an octet operator.
So the allowed mass dependence of the outer ring octet decay constants
is similar to the allowed dependence of the octet masses. Thus we have
\begin{eqnarray}
   f(a\overline{b})
      &=& F_0 + G(\delta m_a + \delta m_b)
                                                         \nonumber  \\
      & & \phantom{F_0}
                   + H_0\sixth(\delta m_u^2 + \delta m_d^2 + \delta m_s^2)
                   + H_1(\delta m_a^2 + \delta m_b^2)
                   + H_2(\delta m_a - \delta m_b)^2
                                                         \nonumber  \\
      & & \phantom{F_0}
                   + \ldots \,.
\label{fpi_NLO_expan}
\end{eqnarray}
The $SU(3)$ flavour symmetric breaking expansion has the simple property
that for any flavour singlet quantity, which we generically denote
by $X_S \equiv X_S(m_u, m_d, m_s)$ then
\begin{eqnarray}
   X_S( \overline{m} + \delta m_u,
                 \overline{m} + \delta m_d,
                 \overline{m} + \delta m_s )
               = X_S(\overline{m}, \overline{m},\overline{m})
                             + O((\delta m_q)^2 ) \,.
\end{eqnarray}
This is already encoded in the above pseudoscalar $SU(3)$ flavour
symmetric breaking expansions, or more generally it can be
shown, \cite{bietenholz10a,bietenholz11a}, that $X_S$ has a
stationary point about the $SU(3)$ flavour symmetric line.

Here we shall consider
\begin{eqnarray}
   X_{\pi}^2 &=& \sixth( M_{K^+}^2 + M_{K^0}^2 + M_{\pi^+}^2 
                           + M_{\pi^-}^2 + M_{\overline{K}^0}^2 + M_{K^-}^2) \,,
                                                           \nonumber \\
   X_{f_\pi} &=& \sixth( f_{K^+} + f_{K^0} + f_{\pi^+}
                            + f_{\pi^-} + f_{\overline{K}^0} + f_{K^-}) \,.
\end{eqnarray}
(The experimental value of $X_\pi$ is $\sim 410\,\mbox{MeV}$, which sets
the unitary range.) There are, of course, many other possibilities
such as $S = N$, $\Lambda$, $\Sigma^*$, $\Delta$, $\rho$, $r_0$,
$t_0$, $w_0$, \cite{bietenholz10a,bietenholz11a,bornyakov15a}.

As a further check, it can be shown that this property also holds
using chiral perturbation theory. For example for mass degenerate
$u$ and $d$ quark masses and assuming $\chi$PT is valid in the region
of the $SU(3)$ flavour symmetric quark mass we find
\begin{eqnarray}
   X_{f_\pi} = f_0\left[ 1 + {8 \over f_0^2}(3L_4 + L_5)\overline{\chi}
                        - 3L(\overline{\chi})
                \right] + O((\delta\chi_l)^2) \,,
\label{XfCPT}
\end{eqnarray}
where the expansion parameter is given by 
$\delta\chi_l = \overline{\chi} - \chi_l$ with
$\overline{\chi} = \textstyle{1 \over 3}(2\chi_l + \chi_s)$,
$\chi_l = B_0m_l$, $\chi_s = B_0m_s$, $f_0$ is the pion
decay constant in the chiral limit, $L_i$ are chiral constants and
$L(\chi) = \chi / (4\pi f_0)^2 \times \ln(\chi / \Lambda_\chi^2)$
is the chiral logarithm. In eq.~(\ref{XfCPT}), as expected,
there is an absence of a linear term $\propto \delta\chi_l$.

The unitary range is rather small so we introduce PQ or partially
quenching (i.e.\ the valence quark masses can be different to the
sea quark masses). This does  not increase the number of expansion
coefficients. Let us denote the valence quark masses by $\mu_q$
and the expansion parameter as $\delta \mu_q  = \mu_q - \overline{m}$.
Then we have
\begin{eqnarray}
   \tilde{M}^2(a\overline{b})
            &=& 1 + \tilde{\alpha}(\delta\mu_a + \delta\mu_b)
                                                   \nonumber   \\
            & &  - (\twothird\tilde{\beta}_1 + \tilde{\beta}_2)
                  (\delta m_u^2 + \delta m_d^2 + \delta m_s^2)
                 + \tilde{\beta}_1(\delta\mu_a^2 + \delta\mu_b^2)
                 + \tilde{\beta}_2(\delta\mu_a - \delta\mu_b)^2 
                                                   \nonumber   \\
            & &  + \ldots \,,
\label{Mpi2twid_NLO_expan}
\end{eqnarray}
and
\begin{eqnarray}
   \tilde{f}(a\overline{b})
            &=& 1 + \tilde{G}(\delta\mu_a + \delta\mu_b)
                                                   \nonumber   \\
            & &
               - (\twothird\tilde{H}_1 + \tilde{H}_2)
                 (\delta m_u^2 + \delta m_d^2 + \delta m_s^2)
                 + \tilde{H}_1(\delta\mu_a^2 + \delta\mu_b^2)
                 + \tilde{H}_2(\delta\mu_a - \delta\mu_b)^2 
                                                   \nonumber   \\
            & &  + \ldots \,,
\label{fpitwid_NLO_expan}
\end{eqnarray}
where in addition to the PQ generalisation we have also formed the ratios
$\tilde{M}^2 = M^2/X_\pi^2$, $\tilde{\alpha} = \alpha/M_0^2$, $\ldots$
and $\tilde{f} = f/X_{f_\pi}$, $\tilde{G} = G/F_0$, $\ldots\,$
(see Appendix~\ref{NNLO_expan} for the NNLO expressions).
This will later prove useful for the numerical results.
We see that there are mixed sea/valence mass terms at NLO (and higher 
orders). The unitary limit is recovered by simply replacing
$\delta\mu_q \to \delta m_q$.


\section{The Lattice}
\label{lattice}


We use an $O(a)$ non-perturbatively improved clover action with
tree level Symanzik glue and mildly stout smeared $2+1$ clover fermions,
\cite{cundy09a}, for $\beta \equiv 10/g_0^2 = 5.40$, $5.50$, $5.65$, $5.80$
(four lattice spacings). We set
\begin{eqnarray}
   \mu_q = {1 \over 2}
           \left( {1\over \kappa_q^{\rm val}} - {1\over \kappa_{0c}} \right) \,,
\end{eqnarray}
giving
\begin{eqnarray}
  \delta \mu_q = \mu_q - \overline{m}
               = {1 \over 2}\left( {1 \over \kappa_q^{\rm val}} 
                       - {1 \over \kappa_0} \right) \,.
\end{eqnarray}
A $\kappa$ value along the $SU(3)$ symmetric line is denoted by $\kappa_0$,
while $\kappa_{0c}$ is the value in the chiral limit. Note that practically
we do not have to determine $\kappa_{0c}$, as it cancels in $\delta\mu_q$.
(For simplicity we have set the lattice spacing to unity.)

\begin{figure}[h]
\begin{center}

\begin{minipage}{0.45\textwidth}

   \begin{center}
      \includegraphics[width=6.00cm]
         {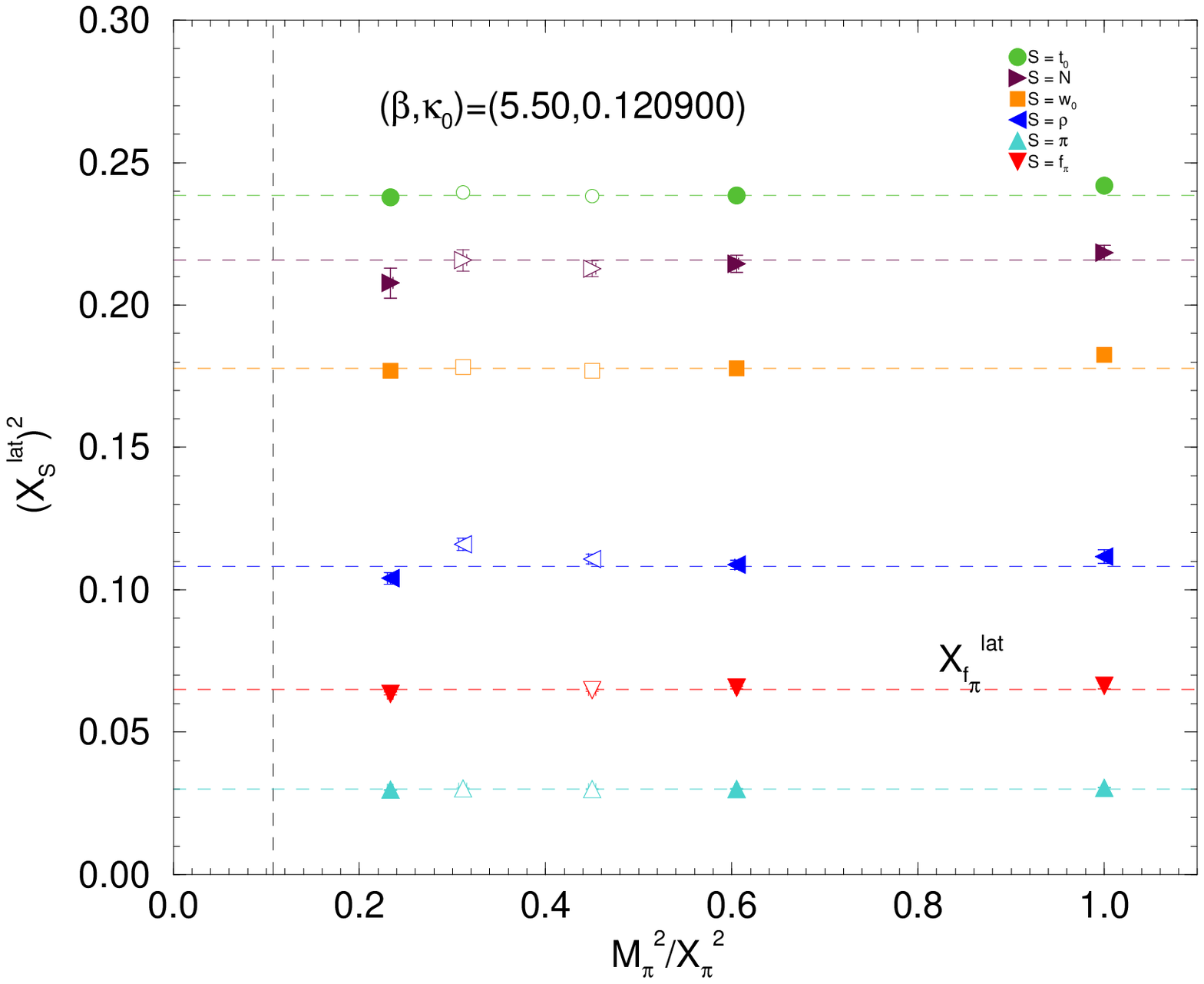}
   \end{center} 

\end{minipage}\hspace*{0.05\textwidth}
\begin{minipage}{0.45\textwidth}

   \begin{center}
      \includegraphics[width=6.00cm]
         {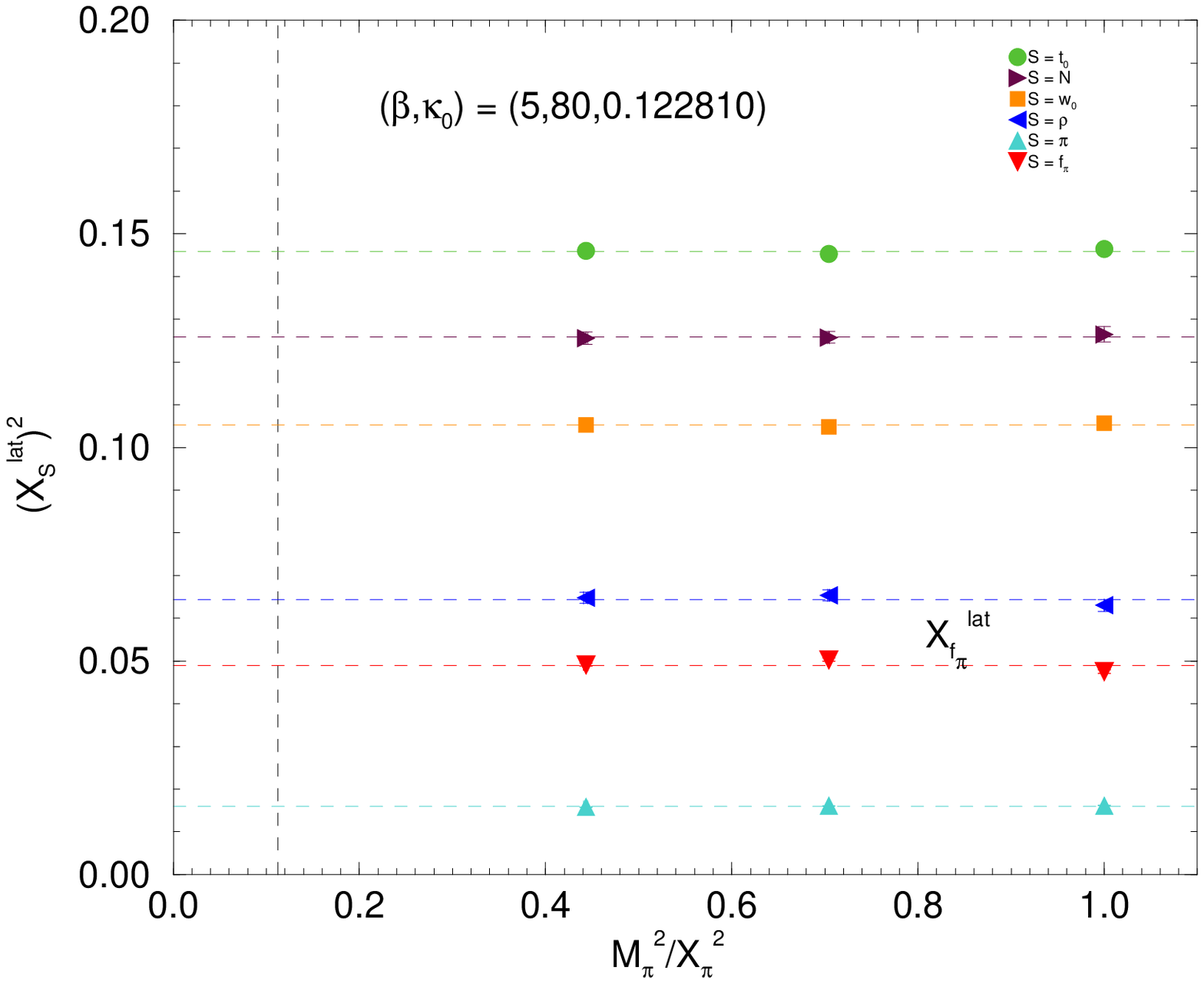}
   \end{center} 

\end{minipage}

\end{center}
\caption{LH panel: $X^2_{t_0}$, $X^2_{w_0}$, $X^2_\pi$, $X^2_\rho$,
         $X^2_N \approx X^2_\Lambda$, $X_{f_\pi}$ for 
         $(\beta, \kappa_0) = (5.50, 0.120900)$
         along the $\overline{m} = \mbox{const.}$ line, together
         with constant fits. Open symbols have $M_\pi L \lsim 4$
         and are not included in the fit. The vertical line is the
         physical point.
         RH panel: The same for 
         $(\beta, \kappa_0) = (5.80, 0.122810)$.}
\label{various_Xs}
\end{figure}

We first investigate the constancy of $X_S$ in the unitary region.
In Fig~\ref{various_Xs} we show various choices for $X_S$.
It is apparent that over a large range, starting from the $SU(3)$
flavour symmetric line, reaching down and approaching the physical point,
$X_S$ appears constant, with very little evidence of curvature. 
(Although not included in the fits, the open symbols have 
$M_\pi L \sim 3$ -- $4$ and also do not show curvature.)
Presently our available pion masses reach down to $\sim 220\,\mbox{MeV}$.

\begin{figure}[h]
\begin{center}

\begin{minipage}{0.45\textwidth}

   \begin{center}
      \includegraphics[width=6.00cm]
         {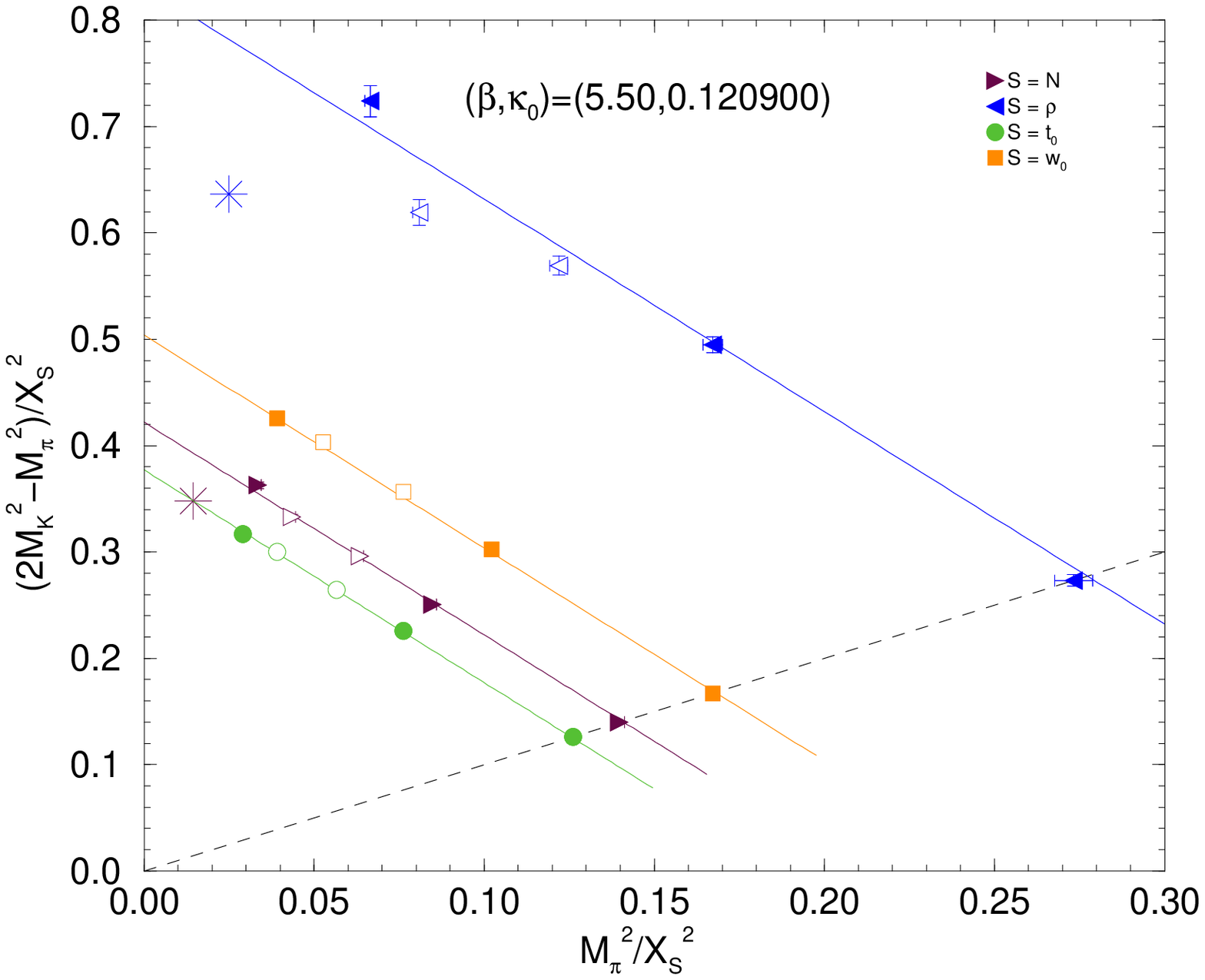}
   \end{center} 

\end{minipage}\hspace*{0.05\textwidth}
\begin{minipage}{0.45\textwidth}

   \begin{center}
      \includegraphics[width=6.00cm]
         {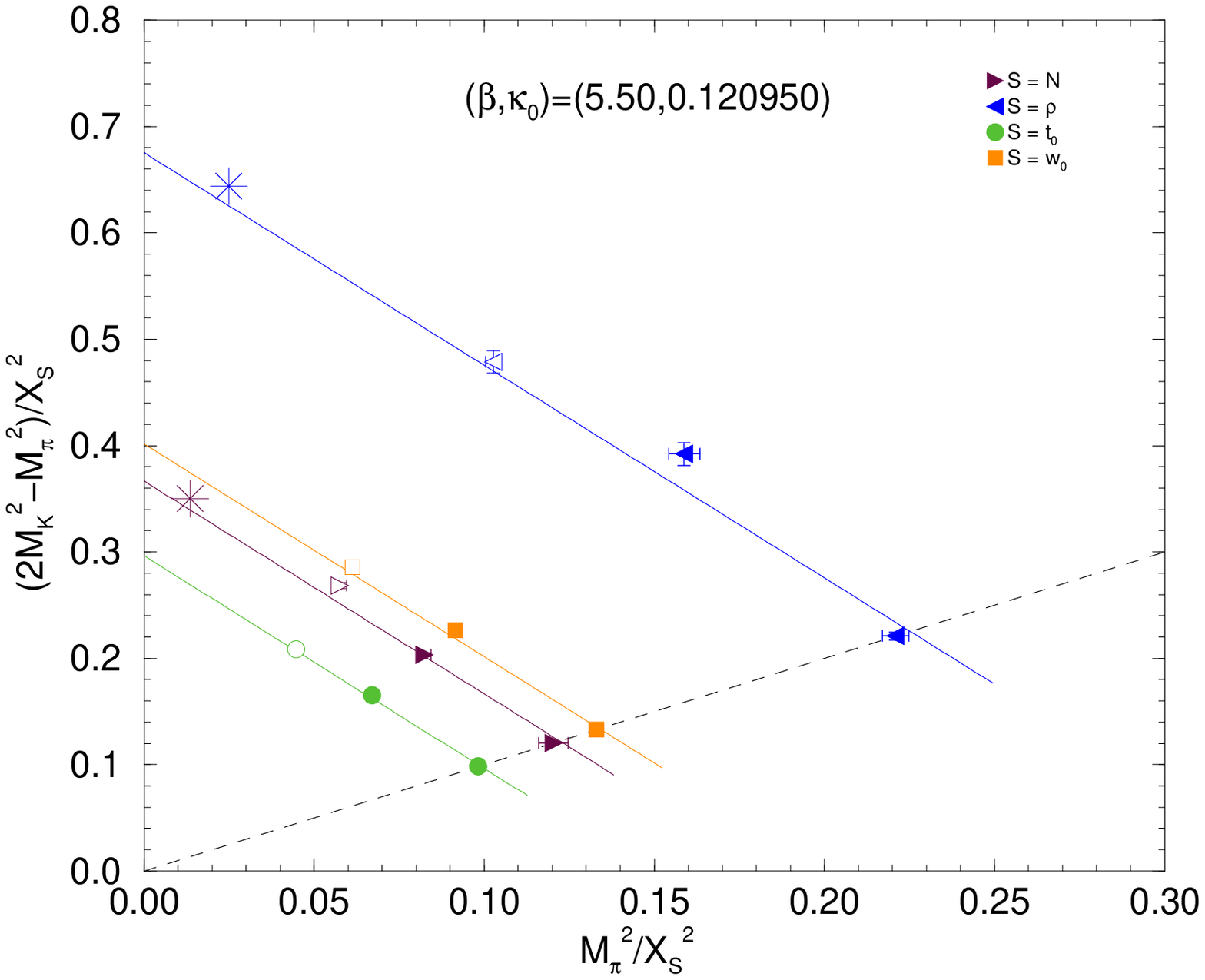}
   \end{center} 

\end{minipage}

\end{center}
\caption{LH panel: $(2M_K^2 - M_\pi^2) / X_S^2$ versus $M_\pi^2 / X_S^2$,
         $S = N$, $\rho$, $t_0$, $w_0$ for
         $(\beta, \kappa_0) = (5.50, 0.120900)$. Stars represent
         the physical points, the dashed line is the $SU(3)$ flavour
         symmetric line.
         RH panel: The same for $(\beta, \kappa_0) = (5.50, 0.120950)$.}
\label{various_paths}
\end{figure}

Based on this observation, we determine the path in the quark mass plane
by considering $M_\pi^2 / X_S^2$ against $(2M_K^2 - M_\pi^2) / X_S^2$.
If there is little curvature then we expect that
\begin{eqnarray}
   {2M_K^2 - M_\pi^2 \over X^2_S} 
      = 3\, {X_\pi^2 \over X_S^2} - 2\, {M_\pi^2 \over X_S^2} 
\end{eqnarray}
holds for $S = N, \rho, t_0, w_0, \ldots$\,. In Fig.~\ref{various_paths}
we show this for $(\beta, \kappa_0) = (5.50, 0.120900)$, $(5.50, 0.120950)$.
We see that this is indeed the case. In addition $\kappa_0$ is adjusted
so that the path goes through (or very close to) the physical value.
For example we see that from the figure, $\beta = 5.50$,
$\kappa_0 = 0.120950$ is very much closer to this path than
$\kappa_0 = 0.120900$, \cite{bornyakov15a}.

\begin{figure}[h]
\begin{center}

\begin{minipage}{0.45\textwidth}

   \begin{center}
      \includegraphics[width=6.00cm]
         {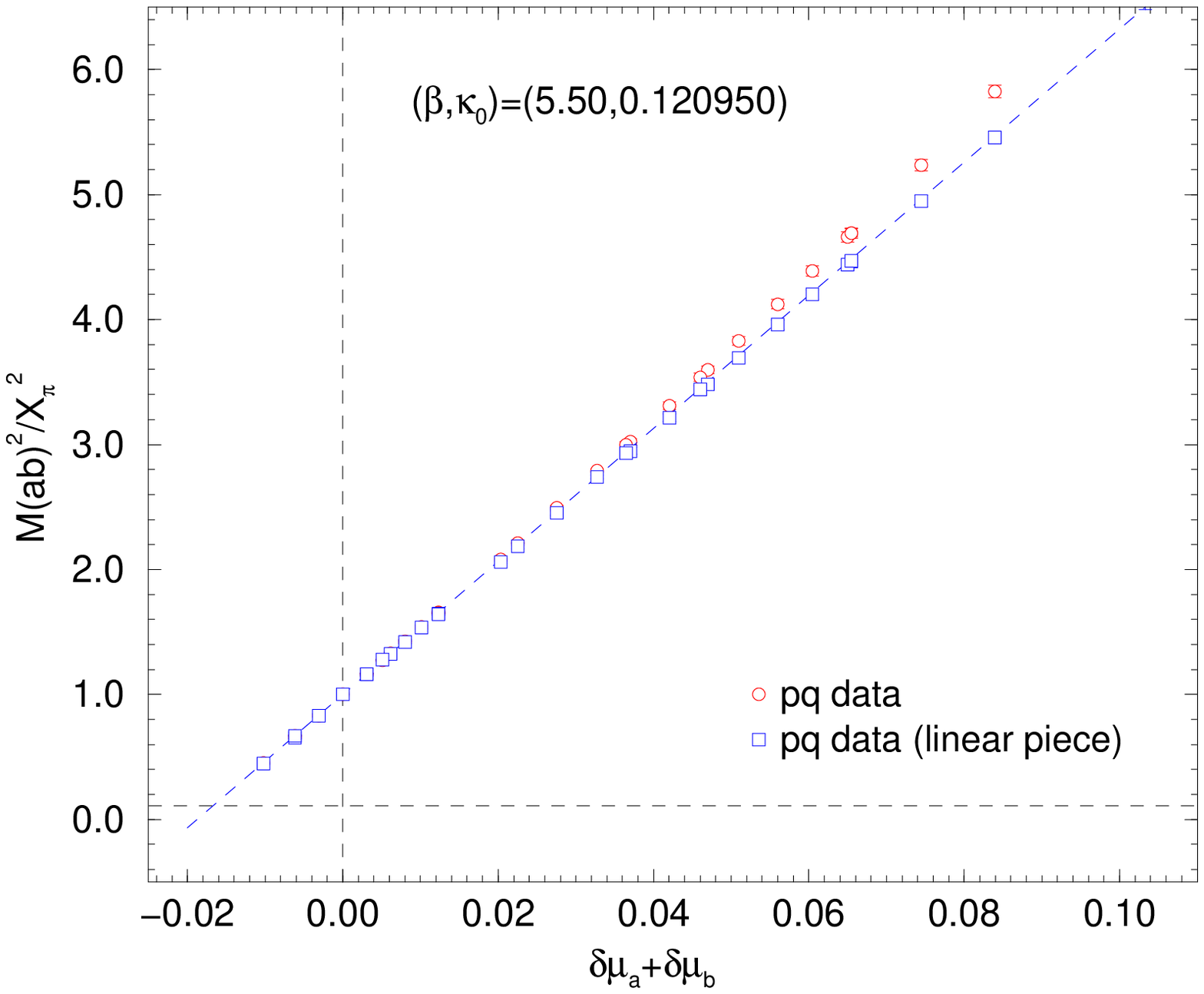}
   \end{center} 

\end{minipage}\hspace*{0.05\textwidth}
\begin{minipage}{0.45\textwidth}

   \begin{center}
      \includegraphics[width=6.00cm]
         {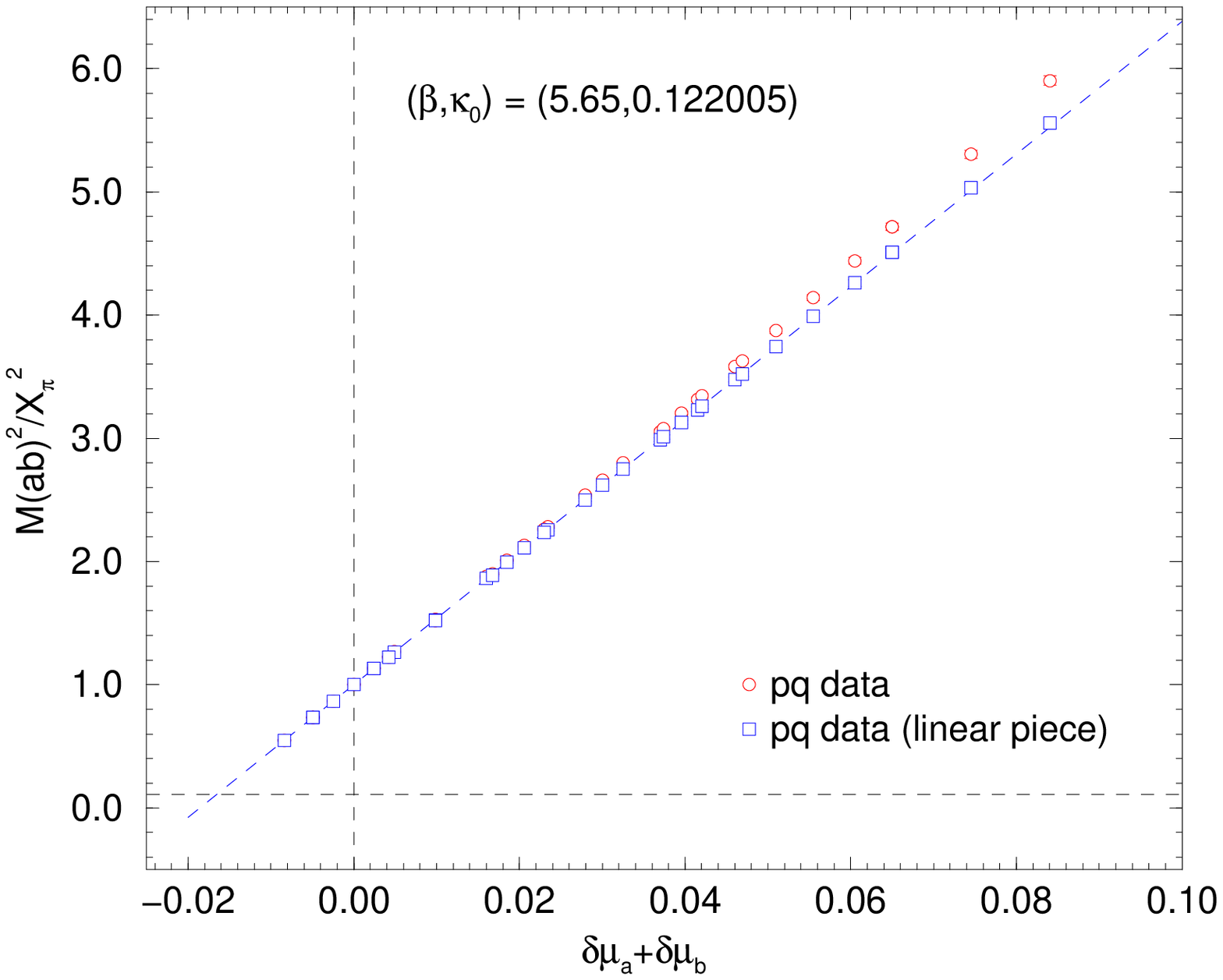}
   \end{center} 

\end{minipage}

\end{center}
\caption{LH panel: PQ (and unitary) pseudoscalar mass results for 
         $\tilde{M}^2 = M^2/X_\pi^2$ with
         $(\beta, \kappa_0) = (5.50, 0.120950)$ against valence
         quarks $\delta\mu_a + \delta\mu_b$. The data is given by
         circles, while subtracting out the non-linear pieces
         (using the fit) gives the squares, together with the
         linear fit. The vertical dashed line is the symmetric point,
         while the horizontal dashed line represents the physical
         $\tilde{M}_\pi^2$. RH panel: Similarly for 
         $(\beta, \kappa_0) = (5.65, 0.122005)$.}
\label{pq_Mtwid2_results}
\end{figure}

\begin{figure}[!htb]
\begin{center}

\begin{minipage}{0.45\textwidth}

   \begin{center}
      \includegraphics[width=6.00cm]
     {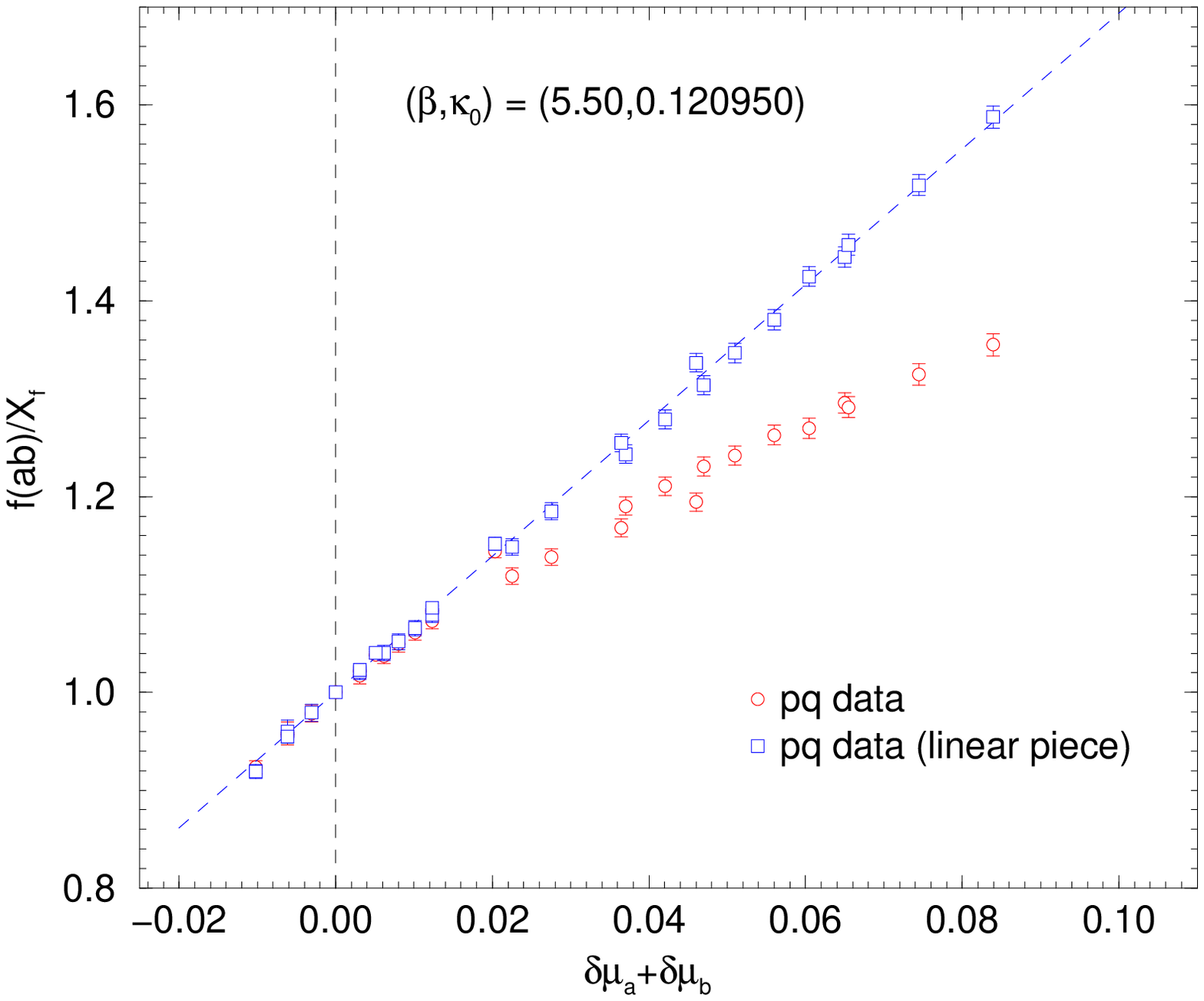}
   \end{center} 

\end{minipage}\hspace*{0.05\textwidth}
\begin{minipage}{0.45\textwidth}

   \begin{center}
      \includegraphics[width=6.00cm]
         {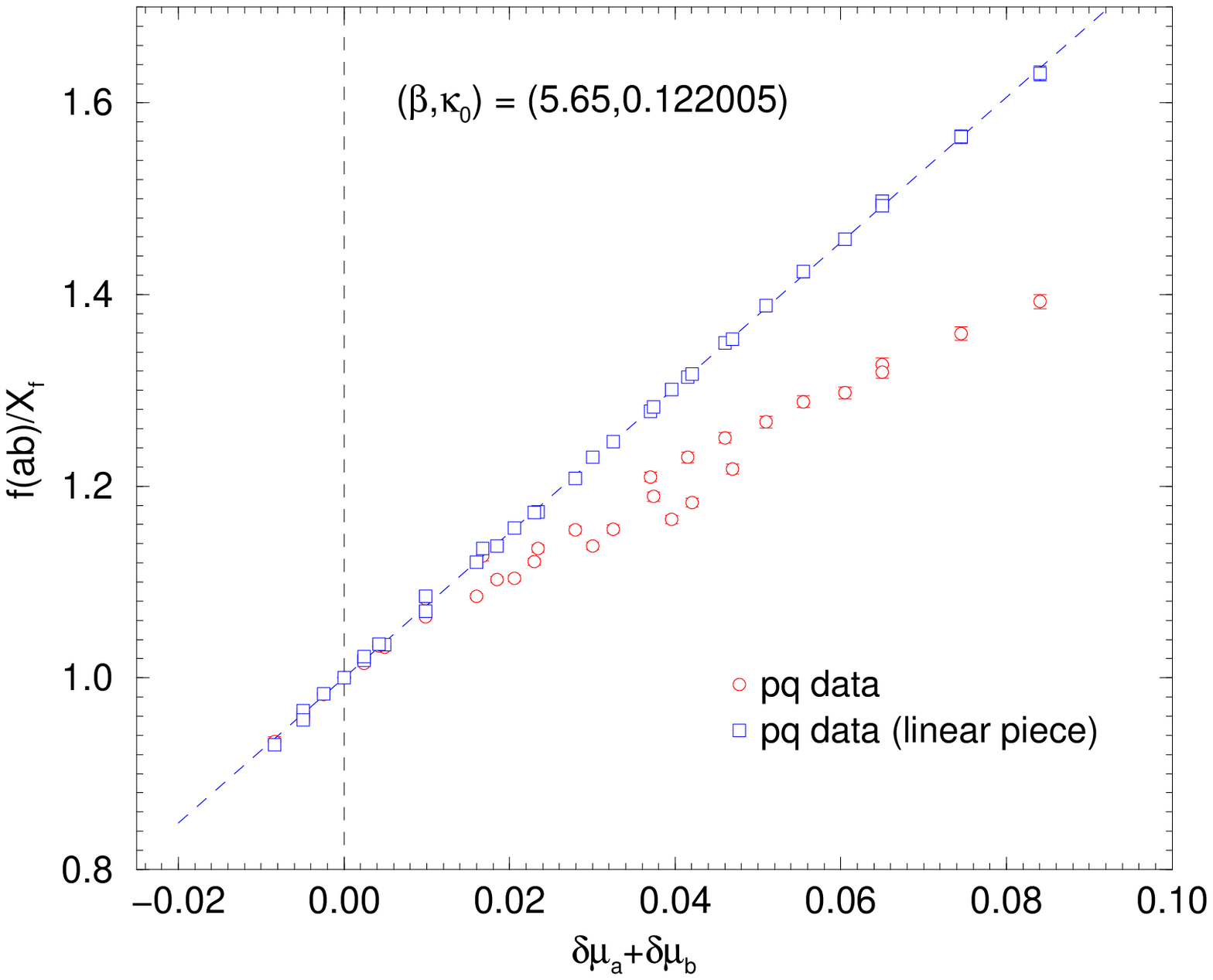}
   \end{center} 

\end{minipage}

\end{center}
\caption{Similarly for the decay constant, $\tilde{f} = f/X_{f_\pi}$.}
\label{pq_ftwid_results}
\end{figure}

The programme is thus first to determine $\kappa_0$ and
then find the expansion coefficients. Then use%
\footnote{Masses are taken from FLAG3, \cite{flag3}.}
isospin symmetric `physical' masses $M_\pi^*$, $M_K^*$ to determine
$\delta m_l^*$ and $\delta m_s^*$. PQ results can help for the first task.
As the range of PQ quark masses that can then be used is much larger
than the unitary range, then the numerical determination of the relevant
expansion coefficients is improved. PQ results were generated about
$\kappa_0$, a single sea background, so $\tilde{\gamma}_1$ was not relevant. 
Also some coefficients (those $\propto (\delta\mu_a-\delta\mu_b)^2$)
often just contributed to noise, so were then ignored.
In Fig.~\ref{pq_Mtwid2_results} we show $\tilde{M}_\pi^2$
against $\delta\mu_a+\delta\mu_b$. From the $SU(3)$ flavour
breaking expansions the leading-order or LO expansions
are just a function of $\delta\mu_a+\delta\mu_b$;
at higher orders, NLO etc.\,, this is not the case
(see eq.~(\ref{Mpi2twid_NLO_expan})).
We see that there is linear behaviour (coincidence of the PQ
data with the linear piece) in the masses at least for 
$\tilde{M}_\pi^2 \lsim 3$ or 
$M_\pi \lsim \sqrt{3}\times 410\,\mbox{MeV} \sim 700\,\mbox{MeV}$.
In Fig.~\ref{pq_ftwid_results} we show the corresponding results for
$\tilde{f}$. Again we see similar results for $\tilde{f}$ as for 
$\tilde{M}^2$; while our fit is describing the data well, the deviations
from linearity occur earlier.

Furthermore the use of PQ results allows for a possibly interesting
method for fine tuning of $\kappa_0$ to be developed. If we
slightly miss the starting point on the $SU(3)$ flavour symmetric line,
we can also tune $\kappa_0$ using PQ results so that we get the
physical values of (say) $M^*_\pi$, $X^*_N$ and $M^*_K$ correct.
This gives $\kappa_0$, $\delta\mu_l^*$, $\delta\mu_s^*$.
The philosophy is that most change is due to a change in valence quark
mass, rather than sea quark mass. Note that then
$2\delta\mu_l^*+\delta\mu_s^* \not= 0$
necessarily (while $2\delta m_l + \delta m_s$ always vanishes).
For our $\kappa_0$ values used here, namely
$(\beta, \kappa_0) = (5.40, 0.119930)$,
$(5.50, 0.120950)$, $(5.65, 0.122005)$, $(5.80, 0.122810)$, 
\cite{bornyakov15a} (on $24^3\times 48$, $32^3\times 64$, 
$32^3\times 64$ and $48^3\times 96$ lattice volumes respectively)
tests show this is a rather small correction and we shall use this
as part of the systematic error, see Appendix~\ref{systematic}.

\begin{figure}[!t]
   \begin{center}

   \begin{minipage}{0.45\textwidth}

   \begin{center}

      \includegraphics[width=6.00cm]
                      {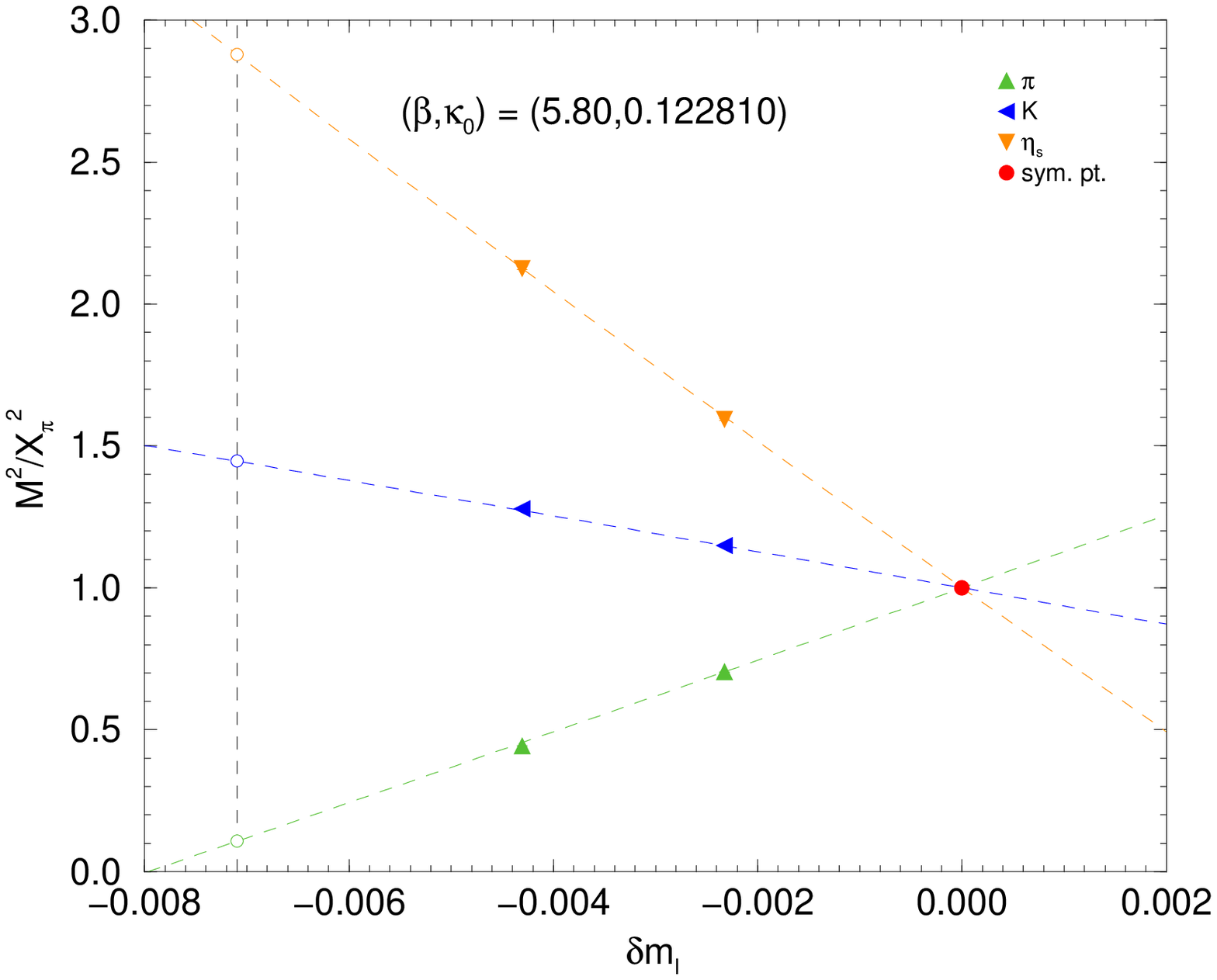}
   \end{center} 

   \end{minipage}\hspace*{0.05\textwidth}
   \begin{minipage}{0.45\textwidth}

   \begin{center}
      \includegraphics[width=6.00cm]
                      {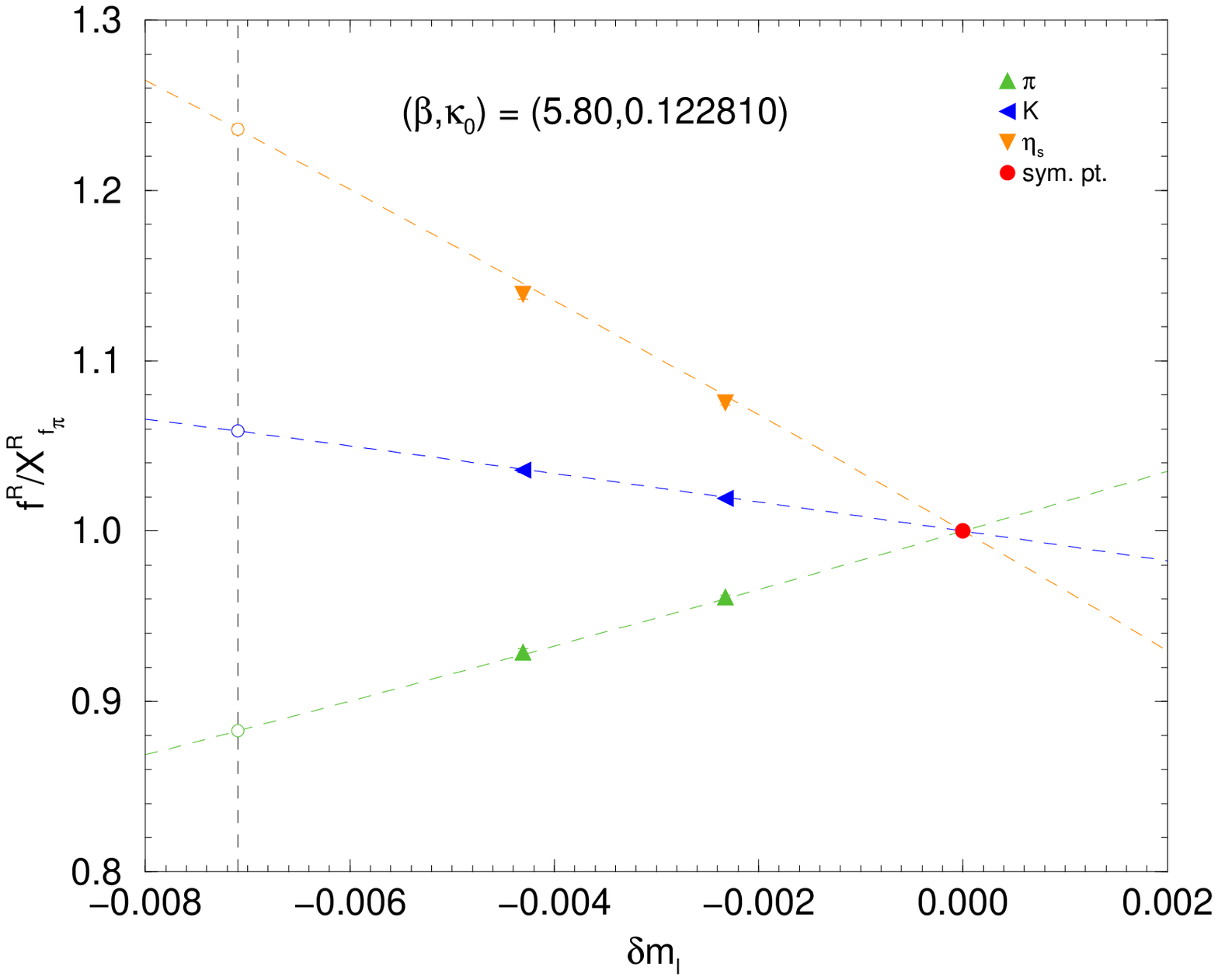}
   \end{center} 

   \end{minipage}

   \end{center}
\caption{LH panel: Unitary results for $\tilde{M}^2 = M^2/X_\pi^2$ versus
         $\delta m_l$ for $(\beta,\kappa_0) = (5.80,0.122810)$.
         RH panel: Equivalent unitary results for $\tilde{f} = f/X_{f_\pi}$.}
\label{fan_plot}
\end{figure}

Of course the unitary range is much smaller, as can be seen
from the horizontal lines in Fig.~\ref{pq_Mtwid2_results}. 
In the LH panel of Fig.~\ref{fan_plot} we show this range as a
function of $\delta m_l$ for $\tilde{M}_\pi^2$, $\tilde{M}_K^2$ and 
$\tilde{M}_{\eta_s}^2$, together with the previously found fits. 
The expressions are given from eq.~(\ref{Mpi2twid_NLO_expan}),
setting $\delta\mu \to \delta m_q$ and then $a \to u$, $b \to d$
with $m_u = m_d \equiv m_l$ for $\tilde{M}^2_\pi$ etc.\,. 
Here we clearly observe the typical `fan' behaviour seen in the
mass of other hadron mass multiplets \cite{bietenholz11a}.
As we have mass degeneracy at the symmetric point, the masses
radiate out from this point to their physical values. For both $\tilde{M}^2$
and $\tilde{f}$ the LO completely dominates.

As can be seen from Fig.~\ref{fan_plot} when $\tilde{M}_\pi$ takes its physical
value, $\tilde{M}_\pi^*$, this determines the physical value $\delta m_l^*$.
These are given in Table~\ref{dmq_phys}.
\begin{table}[!htb]
   \begin{center}
      \begin{tabular}{c|rrrr}
         $\beta$ & \multicolumn{1}{c}{5.40}
                 & \multicolumn{1}{c}{5.50} 
                 & \multicolumn{1}{c}{5.65} 
                 & \multicolumn{1}{c}{5.80}   \\
         \hline
         $\delta m_l^*$  & -0.01041(11)  
                         & -0.008493(33) 
                         & -0.008348(33) 
                         &  -0.007094(11)     \\
      \end{tabular}
   \end{center}
\caption{Results for  $\delta m_l^*$.}
\label{dmq_phys}
\end{table}
Note that due to the constraint given in eq.~(\ref{trivial_constraint})
then $\delta m_s^* = -2\delta m_l^*$.


\section{Decay constants}

\begin{figure}[h]
   \begin{center}

   \begin{minipage}{0.45\textwidth}

   \begin{center}
      \includegraphics[width=6.00cm]{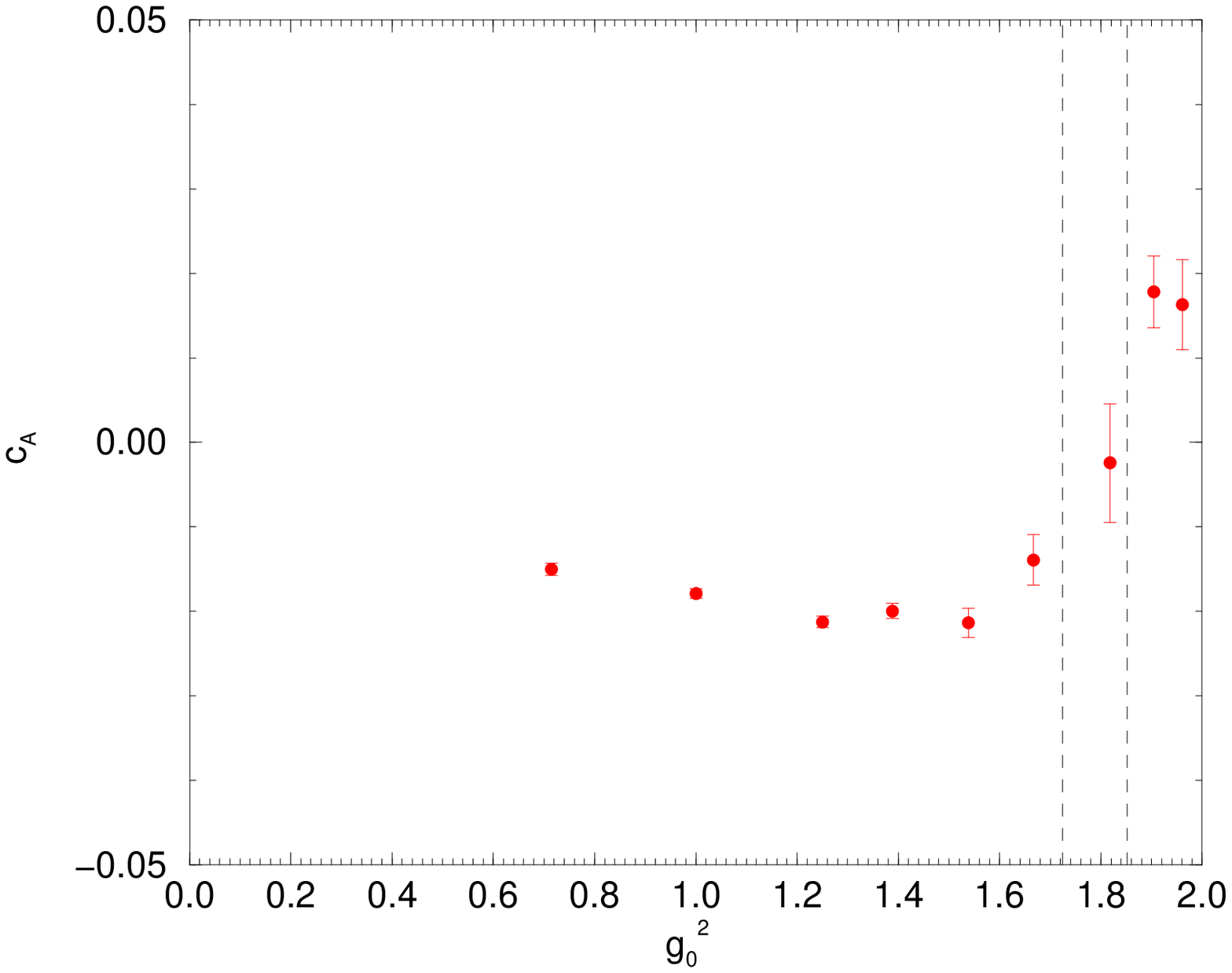}
   \end{center} 

   \end{minipage}\hspace*{0.05\textwidth}
   \begin{minipage}{0.45\textwidth}

   \begin{center}
      \includegraphics[width=6.00cm]
 {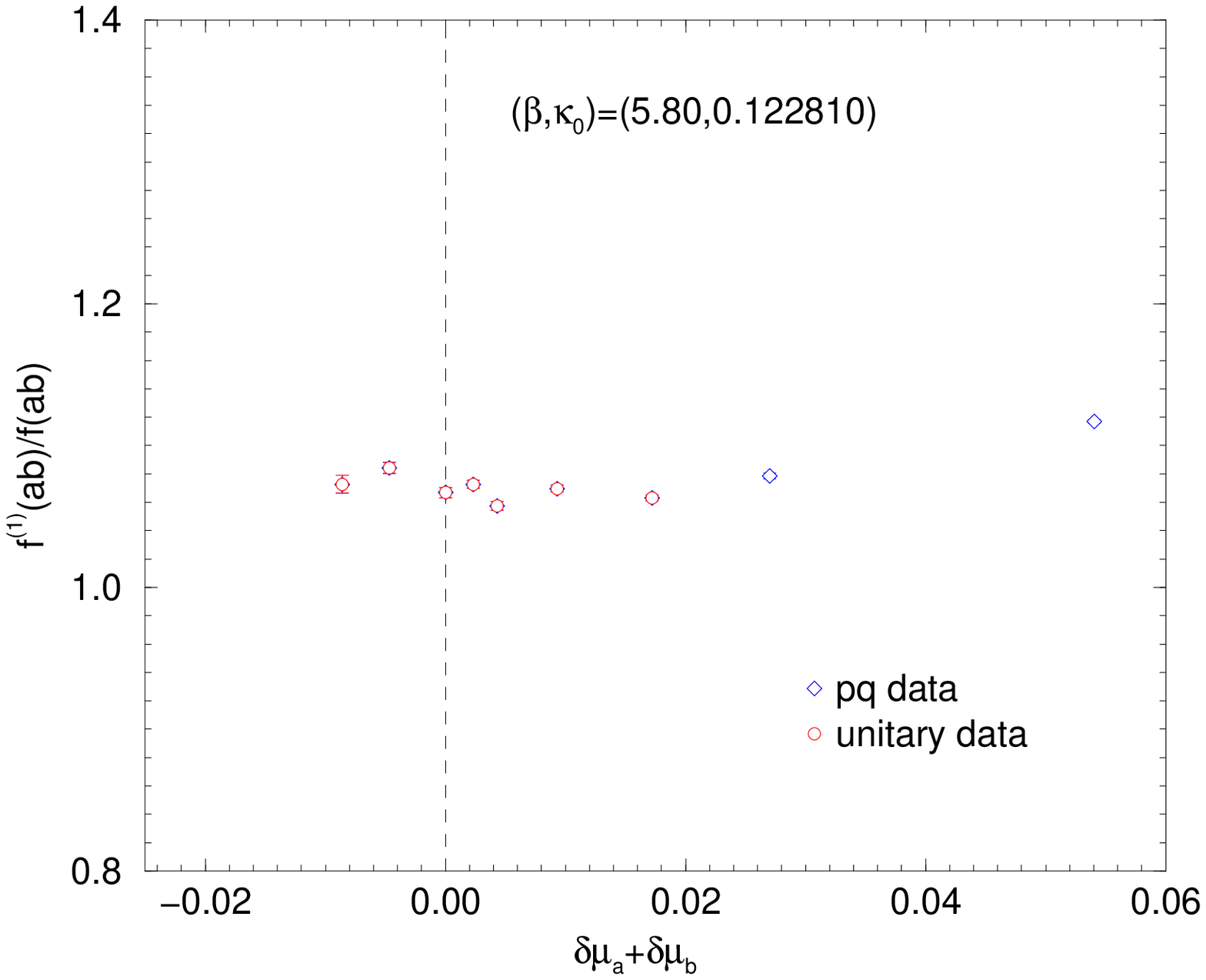}
   \end{center} 

   \end{minipage}

   \end{center}
\caption{LH panel: Estimate of the $c_A$ improvement coefficient
         using the Schr\"odinger Functional,
         \protect\cite{cundy09a} as a function of $g_0^2 = 10/\beta$.
         The vertical dashed lines denote the $\beta$ range
         $5.40$ -- $5.80$. 
         RH panel: The ratio $f^{(1)}/f$ versus $\delta\mu_a+\delta\mu_b$
         for $(\beta,\kappa_0) = (5.80, 0.122810)$.}
\label{improve_stuff}
\end{figure}

The renormalised and $O(a)$ improved axial current is given by
\cite{bhattacharya05a}
\begin{eqnarray}
   {\cal A}^{ab;{\R}}_\mu  = Z_A {\cal A}^{ab;{\imp}}_\mu \,,
\end{eqnarray}
with
\begin{eqnarray}
   {\cal A}^{ab;\imp}_\mu
               = \left( 1 + \left[\overline{b}_A\overline{m} 
                                  +  \half {b}_A (m_a+m_b) \right] 
                 \right) {\cal A}_\mu^{ab}\,, \qquad
   {\cal A}_\mu^{ab}
               = A^{ab}_\mu + c_A\partial_\mu P^{ab} \,,
\end{eqnarray}
and
\begin{eqnarray}
   A^{ab}_\mu = \overline{q}_a\gamma_\mu\gamma_5 q_b\,, \qquad
   P^{ab}    = \overline{q}_a\gamma_5 q_b \,.
\label{Amu+P}
\end{eqnarray}
Using the axial current we first define matrix elements
\begin{eqnarray}
   \langle 0| \widehat{A}_4 | M \rangle = M\, f \,, \qquad
            \langle 0| \widehat{\partial_4 P} | M \rangle = M\, f^{(1)} \,,
\end{eqnarray}
giving for the renormalised pseudoscalar constants
\begin{eqnarray}
   f^{\R} = Z_A\left( 1 + c_A {f^{(1)} \over f} \right)
              \left( 1 + \left[ (\overline{b}_A+b_A)\overline{m}
                           + \half b_A(\delta m_a+\delta m_b)
                     \right]
              \right) f \,.
\end{eqnarray}
As indicated in Fig.~\ref{improve_stuff}, we note that $c_A$ is small
(compared to unity) and that $f^{(1)}/f$ is constant and $\sim O(1)$
in the unitary region. So for constant $\overline{m}$ we can absorb
the $c_A f^{(1)}/f$  and $(\overline{b}_A+b_A)\overline{m}$ terms
to give a change in the first coefficient
\begin{eqnarray}
   \tilde{f}^{\R} \equiv {f^{\R} \over X_{f_\pi}^{\R}}
               = 1 + \left( \tilde{G} + \half b_A \right)
                                    (\delta m_a + \delta m_b) + \ldots \,.
\end{eqnarray}
For $b_A$ (only defined up to terms of $O(a)$) we presently take
the tree level value, $b_A = 1 + O(g_0^2)$.


\section{Results}


\subsection{$f_K / f_\pi$}

\begin{figure}[h]
   \begin{center}
      \includegraphics[width=7.00cm]{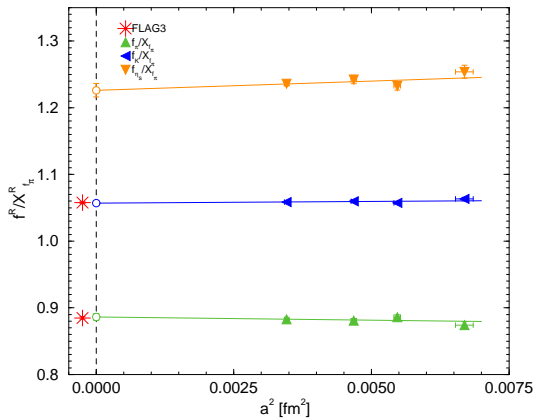}
   \end{center} 
\caption{The continuum extrapolation of $\tilde{f}^{\R*}$. The
         extrapolated values are again given as open circles.
         The converted FLAG3 values,
         \protect\cite{flag3}, are given as stars.}
\label{continuum_results}
\end{figure}

As demonstrated in the RH panel of Fig.~\ref{fan_plot}, we again expect
LO behaviour for $SU(3)$ flavour symmetry breaking for $\tilde{f}$ to
dominate in the unitary region. Using the coefficients
for the $SU(3)$ flavour breaking expansion for $\tilde{f}$
as previously determined, and then extrapolating to the physical quark
masses gives the results in Table~\ref{ftwid_phys}.
\begin{table}[!htb]
   \begin{center}
      \begin{tabular}{cc|rrr}
         $\beta$ & $a\,[\mbox{fm}]$
                 & \multicolumn{1}{c}{$\tilde{f}_\pi^{\R*}$} 
                 & \multicolumn{1}{c}{$\tilde{f}_K^{\R*}$} 
                 & \multicolumn{1}{c}{$\tilde{f}_{\eta_s}^{\R*}$}     \\
         \hline
         5.40    & 0.0818(9) & 0.8739(52)  & 1.0631(26) &  1.2540(97) \\
         5.50    & 0.0740(4) & 0.8859(34)  & 1.0573(17) &  1.2328(63) \\
         5.65    & 0.0684(4) & 0.8806(34)  & 1.0599(17) &  1.2423(62) \\
         5.80    & 0.0588(3) & 0.8827(14)  & 1.0587(07) &  1.2359(28) \\
          \hline
        $\infty$ & 0         & 0.8862(52)  & 1.0568(26) &  1.2263(99) \\
      \end{tabular}
   \end{center}
\caption{Results for $\tilde{f}_\pi^{\R*}$, $\tilde{f}_K^{\R*}$,
         $\tilde{f}_{\eta_s}^{\R*}$, together with the extrapolated
         continuum value.}
\label{ftwid_phys}
\end{table}
Finally using these results, we perform the final continuum 
extrapolation, using the lattice spacings given in \cite{bornyakov15a},
as shown in Fig.~\ref{continuum_results}.
(The fits have $\chi^2/{\rm dof} \sim 3.3/2 \sim 1.6$.)
For comparison, the FLAG3 values, \cite{flag3}, are shown as stars.
(Note that although $f_{\eta_s}$ helps in determining the expansion coefficients,
there is no further information to be found from the various
extrapolated values.) 
Continuum values are also given in Table~\ref{ftwid_phys}.
Converting $\tilde{f}_K^{\R*}$ gives a result of
\begin{eqnarray}
   {f_K \over f_\pi}  =  1.192(10)(13) \,,
\label{fKofpi_result}
\end{eqnarray}
(for simplicity now dropping the superscripts). The first error is
statistical; the second is an estimate of the combined systematic
error due to $b_A$, $SU(3)$ flavour breaking expansion,
finite volume and our chosen path to the physical point
as discussed in Appendix~\ref{systematic}.


\subsection{Isospin breaking effects}


Finally we briefly discuss $SU(2)$ isospin breaking effects.
Provided $\overline{m}$ is kept constant, then the $SU(3)$ flavour
breaking expansion coefficients ($\tilde{\alpha}$,
$\tilde{G}\,,\ldots$) remain unaltered whether we consider
$1+1+1$ or $2+1$ flavours. So although our numerical results
are for mass degenerate $u$ and $d$ quarks we can use them
to discuss isospin breaking effects (ignoring electromagnetic 
corrections). We parameterise these%
\footnote{An alternative, but equivalent method is to first
determine $\delta m_u^*$, $\delta m_d^*$ directly. }
effects by
\begin{eqnarray}
   {f_{K^+} \over f_{\pi^+}} 
               = {f_K \over f_\pi}\left( 1 + \half \delta_{SU(2)} \right) \,,
                                                               \nonumber
\end{eqnarray}
and expanding in $\Delta m = (\delta m_d - \delta m_u)/2$ about the
average light quark mass $\delta m_l = (\delta m_u + \delta m_d)/2$
gives, using the LO expansions (which from Figs.~\ref{pq_Mtwid2_results},
\ref{pq_ftwid_results} or more particularly  Fig.~\ref{fan_plot},
have been shown to work well)
\begin{eqnarray}
   \delta_{SU(2)} 
      = {2 \over 3} \left( 1 - \left({f_K \over f_\pi}\right)^{-1}
                            \right)
                   { \Delta m \over \delta m_l } \,,
\end{eqnarray}
with
\begin{eqnarray}
   { \Delta m \over \delta m_l }
     =  {3 \over 2} \,{ M_{K^0}^2 - M_{K^+}^2 \over
                      M_{\pi^+}^2 - \half\left( M_{K^0}^2+M_{K^+}^2 \right) } \,.
\end{eqnarray}
At the physical point, using the FLAG3, \cite{flag3}, mass values
gives $\Delta m^* / \delta m_l^*$
and hence using our determined value for $f_{K^+} / f_{\pi^+}$,
we find 
\begin{eqnarray}
   \delta_{SU(2)} = -0.0042(2)(2) \,.
\end{eqnarray}
Alternatively, this gives
\begin{eqnarray}
   {f_{K^+} \over f_{\pi^+}} = 1.190(10)(13) \,.
                                                               \nonumber
\end{eqnarray}


\section{Conclusions}


We have extended our programme of tuning the strange and light quark
masses to their physical values simultaneously by keeping
the average quark mass constant from pseudoscalar meson masses
to pseudoscalar decay constants. As for masses we find that
the $SU(3)$ flavour symmetry breaking expansion,
or Gell-Mann--Okubo expansion, works well even at
leading order.

Further developments to reduce error bars could include another
finer lattice spacing, as the extrapolation lever arm in $a^2$ is rather
large and presently contributes substantially to the errors,
and PQ results with sea quark masses not just at the symmetric point
($\kappa_0$) but at other points on the $\overline{m} = \mbox{const}.$ line.


\section*{Acknowledgements}


The numerical configuration generation (using the BQCD lattice
QCD program \cite{nakamura10a}) and data analysis 
(using the Chroma software library \cite{edwards04a}) was carried out
on the IBM BlueGene/Qs using DIRAC 2 resources (EPCC, Edinburgh, UK),
and at NIC (J\"ulich, Germany), the Lomonosov at MSU (Moscow, Russia)
and the SGI ICE 8200 and Cray XC30 at HLRN (The North-German Supercomputer
Alliance) and on the NCI National Facility in Canberra, Australia
(supported by the Australian Commonwealth Government).
HP was supported by DFG Grant No. SCHI 422/10-1 and
GS was supported by DFG Grant No. SCHI 179/8-1.
PELR was supported in part by the STFC under contract ST/G00062X/1
and JMZ was supported by the Australian Research Council Grant
No. FT100100005 and DP140103067. We thank all funding agencies.



\appendix

\section*{Appendix}


\section{Next-to next-to leading order expansion}
\label{NNLO_expan}


We give here the next-to next-to leading order expansion or NNLO expansion
for the octet pseudoscalars and decay constants, which generalise
the results of eqs.~(\ref{Mpi2_NLO_expan}), (\ref{Mpi2twid_NLO_expan}) and
eqs.~(\ref{fpi_NLO_expan}), (\ref{fpitwid_NLO_expan}). For the
pseudoscalar mesons we have
\begin{eqnarray}
   M^2(a\overline{b})
      &=& M^2_0 + \alpha(\delta\mu_a + \delta\mu_b)
                                                         \nonumber   \\
      & & + \beta_0\sixth(\delta m_u^2 + \delta m_d^2 + \delta m_s^2)
                 + \beta_1(\delta\mu_a^2 + \delta\mu_b^2)
                 + \beta_2(\delta\mu_a - \delta\mu_b)^2
                                                         \nonumber   \\
      & & + \gamma_0\delta m_u\delta m_d\delta m_s
                 + \gamma_1(\delta\mu_a + \delta\mu_b)
                              (\delta m_u^2 + \delta m_d^2 + \delta m_s^2)
                                                         \nonumber   \\
      & &        + \gamma_2(\delta\mu_a + \delta\mu_b)^3
                 + \gamma_3(\delta\mu_a + \delta\mu_b)
                              (\delta\mu_a - \delta\mu_b)^2 \,,
\label{Mps2_NNLO_expan}
\end{eqnarray}
and
\begin{eqnarray}
   \tilde{M}^2(a\overline{b})
      &=& 1 + \tilde{\alpha}(\delta\mu_a + \delta\mu_b)
                                                         \nonumber   \\
      & & - (\twothird\tilde{\beta}_1 + \tilde{\beta}_2)
              (\delta m_u^2 + \delta m_d^2 + \delta m_s^2)
                 + \tilde{\beta}_1(\delta\mu_a^2 + \delta\mu_b^2)
                 + \tilde{\beta}_2(\delta\mu_a - \delta\mu_b)^2
                                                         \nonumber   \\
      & & + (2\tilde{\gamma}_2-6\tilde{\gamma}_3)\delta m_u\delta m_d\delta m_s
                 + \tilde{\gamma}_1(\delta\mu_a + \delta\mu_b)
                              (\delta m_u^2 + \delta m_d^2 + \delta m_s^2)
                                                         \nonumber   \\
      & &        + \tilde{\gamma}_2(\delta\mu_a + \delta\mu_b)^3
                 + \tilde{\gamma}_3(\delta\mu_a + \delta\mu_b)
                              (\delta\mu_a - \delta\mu_b)^2 \,.
\label{Mps2twid_NNLO_expan}
\end{eqnarray}
where $\tilde{M}^2(a\overline{b}) = M^2(a\overline{b})/X_\pi^2$
and for an expansion coefficient
$\tilde{\alpha} = \alpha / M_0^2$,
$\tilde{\beta}_i = \beta_i / M_0^2$, $i = 1$, $2$,
and $\tilde{\gamma}_i = \gamma_i / M_0^2$, $i = 1$, $2$, $3$
and we have then redefined $\tilde{\gamma}_1$ by
$\tilde{\gamma}_1 - \tilde{\alpha}(\sixth\tilde{\beta}_0
      +\twothird\tilde{\beta}_1+\tilde{\beta}_2)
                   \to \tilde{\gamma}_1 $.

The $SU(3)$ flavour breaking expansion is identical for the decay
constants, we just replace
$M_0^2 \to F_0$, $\alpha \to G$, $\beta_i \to H_i$,
$\gamma_i \to I_i$ in eq.~(\ref{Mps2_NNLO_expan}) and
$\tilde{\alpha} \to \tilde{G}$, $\tilde{\beta}_i \to \tilde{H}_i$,
$\tilde{\gamma}_i \to \tilde{I}_i$ in eq.~(\ref{Mps2twid_NNLO_expan}).


\section{Correlation functions}
\label{correlation}


On the lattice we extract the pseudoscalar decay constant from 
two-point correlation functions. For large times we expect that
\begin{eqnarray}
   C_{A_4P}(t)
      &=& {1 \over V_S}\, \langle \sum_{\vec{x}}\, A_4(\vec{x},t)
                                  \sum_{\vec{y}}\, P(\vec{y},t) \rangle
                                                          \nonumber  \\
      &=& {1 \over 2M}
           \left[ \langle 0|\widehat{A}_4 |M \rangle
                  \langle 0|\widehat{P} |M \rangle^* e^{-M t} +
                  \langle 0|\widehat{A}^{\dagger}_4 | M \rangle^*
                  \langle 0|\widehat{P}^{\dagger} | M \rangle e^{-M(T-t)}
           \right]
                                                          \nonumber  \\
      &=& - A_{A_4P}
             \left[ e^{-Mt} - e^{-M(T-t)} \right]\,,
\end{eqnarray}
and
\begin{eqnarray}
   C_{PP}(t)
      &=& {1 \over V_S}\, \langle \sum_{\vec{x}}\, P(\vec{x},t)
                                  \sum_{\vec{y}}\, P(\vec{y},t) \rangle
                                                          \nonumber  \\
      &=& {1 \over 2M}
           \left[ \langle 0|\widehat{P} |M \rangle
                  \langle 0|\widehat{P} |M \rangle^* e^{-Mt} +
                  \langle 0|\widehat{P}^{\dagger} |M \rangle^*
                  \langle 0|\widehat{P}^{\dagger} |M \rangle e^{-M(T-t)}
           \right]
                                                          \nonumber  \\
      &=&  A_{PP}
           \left[ e^{-Mt} + e^{-M(T-t)} \right]\,,
\end{eqnarray}
where $A_4$ and $P$ are given in eq.~(\ref{Amu+P}).
We have suppressed the quark indices, so the equations
with appropriate modification are valid for both the pion
and kaon. $V_S$ is the spatial volume and $T$ is the temporal extent
of the lattice. To increase the overlap of the operator with the state
(where possible) the pseudoscalar operator has been smeared using
Jacobi smearing, 
and denoted here with a superscript, $S$ for Smeared.
We now set
\begin{eqnarray}
   \langle 0|\widehat{A}_4|M \rangle
          &=& M f
                                                          \nonumber  \\
   \langle 0|\widehat{\partial_4 P}|M \rangle 
          &=& - \sinh M \langle 0|\widehat{P}|M \rangle
                       = M f^{(1)} \,,
\label{fpi1}
\end{eqnarray}
where $f$, $f^{(1)}$ are real and positive.
By computing $C_{A_4P^{\So}}$ and $C_{P^{\So}P^{\So}}$ we find for
the matrix element
of $\widehat{A}_4$,
\begin{eqnarray}
   M f
      = { \sqrt{2 M} \times {A_{A_4P^{\So}} \over A_{P^{\So}P^{\So}} } 
                     \times \sqrt{A_{P^{\So}P^{\So}}} } \,,
\end{eqnarray}
and for the matrix element of $\widehat{\partial_4P}$
we obtain from the ratio of the $C_{PP^{\So}}$ and $C_{A_4P^{\So}}$
correlation functions 
\begin{eqnarray}
   {f^{(1)} \over f}
        = \sinh M \times { A_{PP^{\So}} \over A_{A_4P^{\So}}} \,.
\label{fpi2}
\end{eqnarray}
Some further details and formulae for other
decay constants are given in \cite{gockeler97a,alikhan07a}.


\section{Systematic errors}
\label{systematic}


We now consider in this Appendix possible sources of systematic errors.


\subsubsection*{Uncertainty in $b_A$}


Presently the improvement coefficient $b_A$ is only known perturbatively
to leading order. We have estimated the uncertainty here by repeating
the analysis with $b_A = 0$ and $b_A = 2$. This leads to a systematic
error on $f_K/f_\pi$ of $\sim 0.008$.


\subsubsection*{$SU(3)$ flavour breaking expansion}


We first note that for the unitary range as illustrated in 
Fig.~\ref{fan_plot}, the `ruler test' indicates there is very little
curvature. This shows that the $SU(3)$ flavour
breaking expansion is highly convergent. (Each order in the
expansion is multiplied by a further power of $|\delta m_l| \sim 0.01$.)
This is also indicated in Fig.~\ref{various_Xs}, where our lowest pion
mass there is $\sim 220\,\mbox{MeV}$. Such expansions are very good compared
to most approaches available to QCD. Comparing the LO (linear)
approximation with the non-linear fit gives an estimation of the
systematic error. The comparison yields the estimate to be
$\sim 0.004$ for $f_K/f_\pi$.


\subsubsection*{Finite lattice volume}


All the results used in the analysis here have $M_\pi L \gsim 4$.
We also have generated some PQ data for $(\beta,\kappa_0) = (5.80,0.122810)$
on a smaller lattice volume -- $32^3\times 64$. (This still has $M_\pi L > 4$.)
Performing the analysis leads to small changes in $\tilde{f}$.
Making a continuum extrapolation (which is most sensitive to just the
$\beta = 5.80$ point) and comparing the result with that of
eq.~(\ref{fKofpi_result}) results in a systematic error of $\sim 0.005$.


\subsubsection*{Path to physical point}


As discussed in section~\ref{lattice}, we can further tune $\kappa_0$
using PQ results to get the physical values $M_\pi^*$, $X_N^*$ and
$M_K^*$ correct, to give $\kappa_0$, $\delta\mu_l^*$, $\delta\mu_s^*$.
Setting $\delta\overline{\mu}^* \equiv (2\delta\mu_l^*+\delta\mu_s^*)/3$
then at LO this average is given by
\begin{eqnarray}
   \delta\overline{\mu}^* 
      = {1 \over 2\tilde{\alpha}}
         \left( \left( \left. {X_\pi^{\rm lat\,2} \over X_N^{\rm lat\,2}} 
                           \right/ {X_\pi^{*2} \over X_N^{*2}}
                \right)^{-1} - 1
           \right) \,,
\end{eqnarray}
(while $2\delta m_l + \delta m_s$ is always $=0$). This gives for
example for $\beta = 5.80$, $ \delta\overline{\mu}^* \sim -0.0001$.
Changing $\delta m_l^*$ (or $\delta m_s^*$) by this and making a continuum
extrapolation (which is again most sensitive to this point) and
comparing the result with that of eq.~(\ref{fKofpi_result}) results in a
systematic error of $\sim 0.009$.


\subsubsection*{Total systematic error}


Including all these systematic errors in quadrature give a total
systematic estimate in $f_K/f_\pi$ of $\sim 0.013$.



\end{document}